\documentclass[manuscript]{aastex}

\usepackage{graphicx,natbib}
\newcommand{\mps}{m\,s$^{-1}$}
\newcommand{\bvec}[1]{ \mbox{\boldmath$#1$} }
\def\id{{\rm d}}

\def\cK{{\cal K}}
\newcommand{\rhobar}{\overline{\rho}}
\newcommand{\vbar}{\overline{v}}
\newcommand{\vv}{\bf v}
\newcommand{\vvb}{\overline{\bf v}}
\newcommand{\dz}[1]{\frac{\partial #1}{\partial z}}

\shorttitle{Verification of helioseismic inversions for surface flows}
\shortauthors{Michal \v{S}vanda}

\begin{document}
\title{Comparison of solar surface flows inferred from time--distance helioseismology and coherent structure tracking using HMI/SDO observations}
\author{Michal \v{S}vanda}
\affil{Astronomical Institute, Academy of Sciences of the Czech Republic (v. v. i.), Fri\v{c}ova 298, CZ-25165 Ond\v{r}ejov, Czech Republic}
\affil{Charles University in Prague, Faculty of Mathematics and Physics, V Hole\v{s}ovi\v{c}k\'ach 2, CZ-18000 Prague 8, Czech Republic}
\email{michal@astronomie.cz}
\author{Thierry Roudier and Michel Rieutord}
\affil{Institut de Recherche en Astrophysique et Plan\'etologie, Universit\'e de Toulouse, 14 avenue Edouard Belin, F-31400, Toulouse, France}
\author{Raymond Burston}
\affil{Max-Planck-Institut f\"ur Sonnensystemforschung, Max-Planck-Stra{\ss}e 2, D-37191 Katlenburg-Lindau, Germany}
\author{Laurent Gizon}
\affil{Max-Planck-Institut f\"ur Sonnensystemforschung, Max-Planck-Stra{\ss}e 2, D-37191 Katlenburg-Lindau, Germany}
\affil{Institut f\"ur Astrophysik, Georg-August-Universit\"at G\"ottingen, D-37077 G\"ottingen, Germany}

\begin{abstract}

We compare measurements of horizontal flows on the surface of the Sun using helioseismic time--distance inversions and coherent structure tracking of solar granules. Tracking provides 2D horizontal flows on the solar surface, whereas the time--distance inversions estimate the full 3-D velocity flows in the shallow near-surface layers. Both techniques use HMI observations as an input. We find good correlations between the various measurements resulting from the two techniques. Further, we find a good agreement between these measurements and the time-averaged Doppler line-of-sight velocity, and also perform sanity checks on the vertical flow that resulted from the 3-D time--distance inversion.

\end{abstract}


\section{Motivation}

The Sun is a very dynamic system, where plasma flows play a very important role in solar dynamo processes. The solar dynamo changes the solar magnetic fields and gives rise to all phenomena of solar activity. Thus the measurement of plasma flows is an important part in understanding the physics of the Sun. Plasma flows in the solar photosphere were studied by many authors by many methods in the past, some methods are straightforward and intuitive whereas some are more complex. Although there is general agreement on the multiscale structure of these flows, details still differ with the method used to infer these flows. It is not easy to answer the question, which method gives the most accurate results. However, by directly comparing various methods one can gain confidence in the trustworthiness of the results.

\subsection{Comparisons between methods}

Not counting the direct measurement of the line-of-sight component of the three-dimensional plasma flow vector, there are two principal methods for measuring velocity fields on the solar surface. The first is based on following structures (granules, supergranules, magnetic elements, etc.) carried by the underlying flow in the series of frames capturing the same region on the Sun. These provide in general two (horizontal) components of the flow. The second (helioseismology) is based on the analysis of the propagation of solar waves through the convective envelope and can, in principle, provide us with all three components of the flow velocity vector.

Both groups of methods have been carefully tested, usually by utilising known inputs from numerical simulations \cite[e.g.][]{RRLNS01} or by comparing applications of the methods to data from different sources. \cite{2006AA...458..301S} validated and calibrated the Local Correlation Tracking (LCT) method \citep[][]{1986ApOpt..25..392N} applied to Michelson Doppler Imager (MDI) full-disc Dopplergrams. Recently \cite{Roudier2013} applied their Coherent Structure Tracking (CST) code to both full-disc Helioseismic and Magnetic Imager (HMI) and Hinode intensitygrams. The agreement between the results inferred from completely different sources was great, except for a systematic error in lower-resolution SDO/HMI velocity maps near the solar limb. A correction curve was derived based on this comparison.

Helioseismic methods were validated in a similar manner. \cite{2007ApJ...657.1157G} applied time--distance \citep{1993Natur.362..430D} methods to a numerical simulation of near-surface convection, where the solar waves are naturally excited. They used only the surface gravity ($f$) mode to derive the horizontal surface flow field by means of scaling the travel-time maps (thus no inversion was performed; such approach is acceptable for the $f$-mode when the surface flow is in question). The comparison of the horizontal ($v_x$ and $v_y$ components) flow maps to the known flows in the simulation ended up with correlation coefficients highly positive (0.70 for $v_x$ and 0.73 for $v_y$). At the same time they also applied LCT to the series of simulated intensitygrams with a correlation to the known flows of 0.99 for both components. However, the authors did not mention a direct comparison of time--distance and LCT velocity maps. \cite{2007ApJ...659..848Z} used the same numerical simulation as \cite{2007ApJ...657.1157G} to validate a proper time--distance inversion for flows with a correlation of the resulting surface flow maps with the known flows in the simulation of 0.72 for $v_x$, 0.64 for $v_y$, and $-0.72$ for vertical $v_z$. Note especially the significant anti-correlation in the vertical flow inversion. It was discussed that this is due to the presence of a cross-talk (i.e. leakage of the horizontal components into the vertical one in the mass-conserving flow).

This conclusion was largely confirmed by \cite{Svanda2011}, who validated a brand-new time--distance inversion code against synthetic data coming from a numerical simulation of Sun-like convection. Only the inverse modelling part of the whole processing pipeline was validated, which allowed avoiding possible problems in the travel-time measurements and to study all components of the inversion in detail. It turned out that, indeed, for weak vertical flow inversions the cross-talk contributions from horizontal components are larger than the proper signal of the vertical flow. The cross-talk component was also highly anti-correlated, which explains the negative sign noted by \cite{2007ApJ...659..848Z}. A method of minimisation of such cross-talks was invented and successfully implemented. The flow estimates from the inversion compared well to the input flows from the numerical simulation with a correlation coefficient of 0.9 for horizontal components in the near-surface layers and of 0.8 for the vertical component (before the minimisation of the cross-talk it was $-0.63$).

Recently, similar end-to-end validation of the inversion pipeline for helioseismic holography (including the measurement of helioseismic observables) was performed by \cite{2013SoPh..282..361D}. The validation was successful in terms of horizontal components and failed (again, due to the cross-talk contributions) for the vertical component.

Comparisons between different methods for retrieving velocity fields were also done in the past. \cite{2004ApJ...613.1253H} compared the results of two distinct helioseismic methods, ring-diagram and time--distance inversions, with high a correlation ($\sim$0.8) between the different flow estimates. A general conclusion was that the two distinct helioseismic methods observe the same Sun. \cite{2000SoPh..192..351D} compared the horizontal divergence of the flow computed from the horizontal velocity maps measured by means of LCT and a proxy for flow divergence from time--distance $f$-mode travel-time maps. The correlation of the two methods was 0.89 when the different resolutions of the horizontal divergence estimates was taken into account. Inconclusive results were obtained by \cite{2005ASPC..346....3A}, using flow maps derived from a series of low-resolution full-disc magnetograms measured, by means of LCT when compared to down-sampled velocity maps from time--distance helioseismology. The correlation coefficient describing the match of the velocity maps resulted from the two methods was close to zero, however, there were compact and continuous regions of characteristic size from 30 to 60 heliographic degrees with a good agreement between the two methods, so that one could not conclude that the results were completely different. \cite{2007SoPh..241...27S} directly compared surface flow maps obtained using LCT applied to SOHO/MDI full-disc Dopplergrams and using time--distance helioseismology. A map-to-map correlation coefficient was 0.82 for the component of the flow in the direction of solar rotation (which was not removed and thus the correlation was dominated by rotation) and 0.58 for component in the south--north direction. The effective resolution of maps compared was set by the size of the LCT apodization window and was around 44~Mm.

To our knowledge, the direct validation of the time--distance inversion pipeline (including the travel-time measurements) and tracking measurements on smaller-than-supergranular scales has not been published. However, it is an important task and must be accomplished when the robustness and trustworthiness of helioseismic flow maps comes in question. The aim of this study is to perform end-to-end testing of the time--distance inversion pipeline running at Max-Planck-Institut f\"ur Sonnensystemforschung (MPS) in Katlenburg-Lindau, Germany and newly implemented also at Astronomical Institute of Academy of Sciences in Ond\v{r}ejov, Czech Republic. By testing the measurements of surface flow we inherently shed more confidence on our knowledge of the flow below the surface and measured by the time--distance helioseismology, which certainly has consequences to the physics of these flows.

\section{Measured surface flows}

We measured surface flow fields using the two distinct methods (tracking and time--distance inversion) applied to data series observed by Helioseismic and Magnetic Imager \citep[HMI;][]{2012SoPh..275..207S,2012SoPh..275..229S} aboard the Solar Dynamics Observatory satellite on the 12th and 13th of May, 2010. The results were compared to test the correctness of the helioseismic inverse modelling.

\subsection{Flow estimate from granulation tracking}

One implementation of tracking methods has a form of \emph{coherent structure tracking} \cite[CST;][]{2007AA...471..687R}. The new implementation of the algorithm \citep{Roudier2012} splits into three main steps:

\begin{enumerate}

\item In each frame of the series of intensitygrams, individual granules are segmented out. The segmentation is based on localisation of local maxima of intensity through the intensity curvature, which are tagged as granular centres. The edges or granular cells are detected by extension of regions around the detected centres with points whose minimal curvature value (evaluated in eight directions from surrounding pixels) is greater than a given threshold, while keeping the minimal distance of one pixel between each pair of granules. In this way, there is a control over the size of segmented structures. Each granule is then identified and tagged.

\item For each granule, a trajectory throughout the data series is drawn by cross-identifying the granules in consecutive frames of data series. Strong assumptions apply in this step. Granules are cross-identified taking into account a common granule area in the consecutive frames of the series with a threshold that is determined by an upper bound on velocity (measured as displacement divided by time lag between the frames). Typically, velocities larger than 7~k\mps{} are rejected. The detected trajectories are usually averaged over a time window, the average velocity is then computed as total displacement change divided by the length of the time window.

\item It is clear that the velocity field derived in the previous step is irregularly spaced. For further analysis, it is necessary to approximate the velocity field by the best differentiable field. That is done using multi-resolution analysis \citep[e.g.][]{1989ITPAM..11..674M}, where the irregular velocity field is expressed in wavelet components with different scales. The differentiable field is then reconstructed from these components.

\end{enumerate}

The CST algorithm has been utilised in many papers \citep[e.g.][to name a few]{2007AA...471..695T,2008AA...479L..17R,2010AA...512A...4R,Roudier2012} and has proved its usefulness in studies of surface velocity fields.

We applied the code to a series of HMI full-disc intensitygrams. The implementation of the CST algorithm used in this study is not sensitive to signal of oscillations in intensity images, thus it is not filtered out. The total data series spanned two full days and this was split into shorter series to capture the evolution of surface flows. Each map was computed with a 30 minute time window, thus we computed 96 maps of the horizontal components of the flow over the studied two days. The results computed using the full-resolution of HMI were binned seven times to increase signal-to-noise ratio. The random-error estimate is 250~m/s for each 30-minutes map \citep{Roudier2012}. In these maps, motions within supergranules visually dominate the flow field.

% Random error estimate: take one map and the other after 30 minutes. The principal components should be supergranules, thus should be more-or-less stable in this time span. So any difference is due to the realisation of noise. RMS of the difference is 300~\mps{} for both $v_x$ and $v_y$. In the 12-hour cube 24 maps are averaged, thus the realisation noise should decrease as $\sqrt(T)$. Thus, this gives an estimate of noise level in the tracked velocities 60~\mps.

\subsection{Flow estimate from time--distance inversion}

The saying ``someone's noise is someone else's signal'' is especially true for helioseismologists \citep[for a recent review, see][]{2010arXiv1001.0930G}. The time series of observations contains not only the convective motions but also solar oscillations. In the previous methods, the signal of solar oscillations did not carry any useful information that could be used by the CST algorithm, and could thus be easily removed without having effect on the analysis. However, the solar oscillations observed in surface measurements are manifestations of seismic waves travelling through the solar interior, where their propagation is affected by anomalies in plasma state parameters. The deviations in travel times of solar waves can be measured and used in order to dig out the information about these anomalies. Here we focus on the inference of flows, which have the largest impact on \emph{difference} travel times, i.e. the difference between the measured travel times of waves travelling in opposite directions.

Time--distance helioseismology \citep{1993Natur.362..430D} is a set of tools useful for measuring and interpreting wave travel times. It consists of the following steps:
\begin{enumerate}

\item Wave travel-times are at best measured from a series of full-disc maps of line-of-sight velocity (Dopplergrams). HMI provides such measurements with high cadence (45~s) thus providing us with almost ideal datasets. The datasets are tracked and remapped to Postel projection using standard data processing techniques. We track only the disc-centre region (512$\times$512 pixels) with a pixel size of 1.46~Mm always for 12 hours. Thus we obtained four consecutive 12-hour Dopplergram datacubes suitable for travel-time measurements (see Tab.~\ref{tab:cubes}). We focus on a small patch near the disc centre, thus Postel projection approximates the Cartesian coordinate system $(x,y,z)$, where $x$ is in the direction of solar rotation, $y$ in direction solar south to north and $z$ indicates height. In the small field-of-view, the deviations of Postel-projected coordinates from the Cartesian system are negligible.

\item The datacubes were spatio-temporarily filtered to separate different modes of solar waves. In this study we use only the surface gravity ($f$) mode.

\item The travel-times were measured following the approach of \cite{2004ApJ...614..472G}. It provides linearised fitting of travel-times from measured signal cross-covariances and is robust in the presence of noise. Travel-times were measured for centre-to-annulus and centre-to-quadrant geometries \citep{1997SoPh..170...63D} with radii of the annuli 5 to 20 pixels, thus providing a set of 48 travel-time maps for each of four Dopplergram datacubes.

\item These travel-times are inverted for flows using inversion weights, which is described separately.
\end{enumerate}

\subsubsection{Time--distance inversion}

We assume that there is a linear relationship between the measured travel-time deviations $\delta\tau$ and velocity vector $\bvec{v}$ given by
\begin{equation}
\delta\tau^a(\bvec{r}) = \int_{\odot} \bvec{K}^a(\bvec{r'}-\bvec{r},z) \cdot \bvec{v}(\bvec{r'},z)\; \id^2\bvec{r'} \, \id z + n^a(\bvec{r})\ .
\label{eq:traveltimesdef}
\end{equation}
Here $\bvec{K}^a = (K_x^a, K_y^a, K_z^a)$ is a vector travel-time sensitivity kernel computed in the Born approximation \citep{2007AN....328..228B}, $\bvec{r}=(x,y)$ is a horizontal position vector, and index $a$ uniquely refers to a particular geometry of the travel-time measurement.

We attempt to invert for $\bvec{v}$ from (\ref{eq:traveltimesdef}) when knowing $\delta\tau$ and $\bvec{K}$. A realisation of the random noise $n^a$ is not known, however its covariance matrix was measured from the data using the $1/T$ fitting approach \citep[for details see][]{2004ApJ...614..472G}, where $T$ is the length of observation. The inversion can be solved by means of a Subtractive Optimally Localised Averaging \citep[SOLA;][]{1992AA...262L..33P} approach. The SOLA technique is standard in time--distance helioseismology \citep[e.g.][]{2008SoPh..251..381J}. It aims to construct a spatially bound averaging kernel by linearly combining the set of sensitivity kernels while keeping the error magnification under control. The estimate for the flow component $v_\alpha^{\rm inv}$, where $\alpha=x,y,z$, is then given by
\begin{equation}
v_\alpha^{\rm inv}(\bvec{r}_0,z_0)=\sum_a \int w_a^\alpha (\bvec{r'}-\bvec{r}_0;z_0) \tau^a(\bvec{r'}) \id^2\bvec{r'},
\label{eq:inversion}
\end{equation}
where $w_a^\alpha$ are the inversion weights to be determined.

The inverted flow velocity is then a combination of (1) a true velocity component smoothed by the averaging kernel $\cK^\alpha_\alpha$, (2) a crosstalk from other components, which we attempt to minimise, and (3) a random-noise component $v_\alpha^{\rm noise}$, root-mean-square value of which ($\sigma_\alpha$) we bound:
\begin{equation}
v_\alpha^{\rm inv}(\bvec{r}_0,z_0)=\int_\odot \cK^\alpha_\alpha(\bvec{r}-\bvec{r}_0,z;z_0) v_\alpha(\bvec{r},z) \id^2\bvec{r} \id z+\sum_{\beta \ne \alpha} \int_\odot \cK^\alpha_\beta(\bvec{r}-\bvec{r}_0,z;z_0) v_\beta(\bvec{r},z) \id^2\bvec{r} \id z+v_\alpha^{\rm noise}(\bvec{r}_0,z_0).
\end{equation}

In terms of the weights, the component of the averaging kernel $\cK_\beta^\alpha$ is expressed by
\begin{equation}
\cK_\beta^\alpha(\bvec{r}, z;z_0)=\sum_a \int_\odot w_a^\alpha(\bvec{r'};z_0) K_\beta^a(\bvec{r}-\bvec{r'},z) \id^2\bvec{r'}.
\end{equation}

The computation of the inversion weights is a mathematical problem, which usually ends up in a need to invert a large ill-posed and ill-conditioned matrix. We use the multichannel approach \citep{fastOLA} assuming the spatial invariance of the background. The multichannel formulation reduces the computational demands significantly.

\begin{figure}
\includegraphics[width=0.49\textwidth]{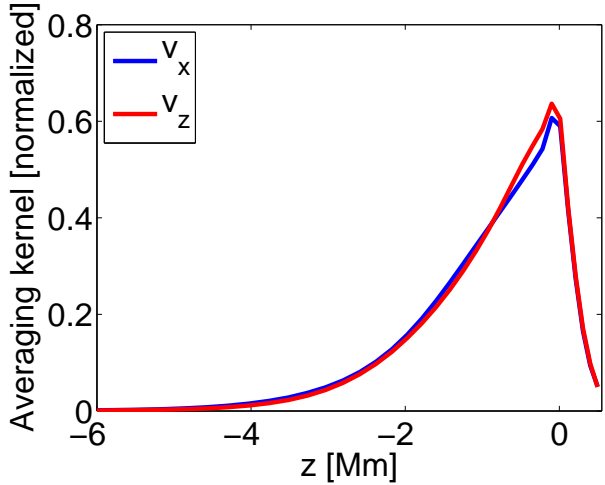}
\caption{Cuts through averaging kernel of inversion for horizontal flow ($v_x$, identical for $v_y$) and vertical flow ($v_z$) as a function of depth.}
\label{fig:akerns}
\end{figure}

We used the inversion code that was already validated using synthetic data \citep{Svanda2011}. The inversion we performed was one of the simplest ones. In accordance with the travel-time measurements described above, we utilised sensitivity kernels $\bvec{K}^a$ for the $f$ mode, where $a$ refers to one of the 48 possibilities combining one of three centre-to-annulus and centre-to-quadrant geometries and one of sixteen radii of the annuli. The exception was only the inversion for the vertical flow, where we used only the centre-to-annulus geometry with the combination of 16 distances. The other two geometries, sensitive to waves travelling in the east--west and south--north directions do not contain much information about the vertical flow, do not contribute to the averaging kernels and just increase the level of noise. We set a requirement on the averaging kernel so that its component in the direction of inversion $\cK^\alpha_\alpha$ is strongly localised near the surface (see Fig.~\ref{fig:akerns}) with a horizontal Full-Width-at-Half-Maximum $FWHM_h=10$~Mm and cross-talk components $\cK^\alpha_\beta$ for $\beta \ne \alpha$ are minimised \citep[for details how to do so we refer to][]{Svanda2011}. The bounds on root-mean-square value of the random-noise component $v_\alpha^{\rm noise}$ are set to be 25~\mps{} and 4~\mps{} for horizontal and vertical components respectively when assuming the averaging over 12 hours.

\subsection{Alignment of flow maps from the two methods}

Unfortunately, although the data entering the two analyses originate from the same telescope and are thus perfectly co-aligned in the beginning, in the course of processing, the coordinate system was changed and the resulting flow maps had to be carefully co-aligned again.

The 2-D flow maps from CST were re-projected onto coordinates grid of the helioseismic datacubes utilising Postel projection with pixel size of 1.46~Mm, and averaged in time to match time--distance time span. For each of four datacubes, 24 CST flow maps were averaged to form one 12-hour average. Averaging over 12 hours increases the signal-to-noise ratio again. Assuming that the realisation noise decreases as $\sqrt{T}$, where $T$ is the length of the time window, the estimate of the noise level in the 12-hour velocities is 50~\mps.

Exact coordinates of the projection centres and time span are given in Table~\ref{tab:cubes}. The angular differential rotation is of the form

\begin{equation}
\omega(b)=2.6373-0.3441\sin^2b-0.5037\sin^4b ~ \mu{\rm rad/s},
\end{equation}
where $b$ stands for the heliographic latitude, was subtracted from re-projected $v_x$ component. Using this rate, the datacubes for time--distance analysis were tracked, thus it is natural to remove precisely this rotation profile from the non-tracked CST flow estimate.

\section{Comparison of flow estimates}
\label{sect:cst_vs_td}

\begin{figure}
\includegraphics[width=0.9\textwidth]{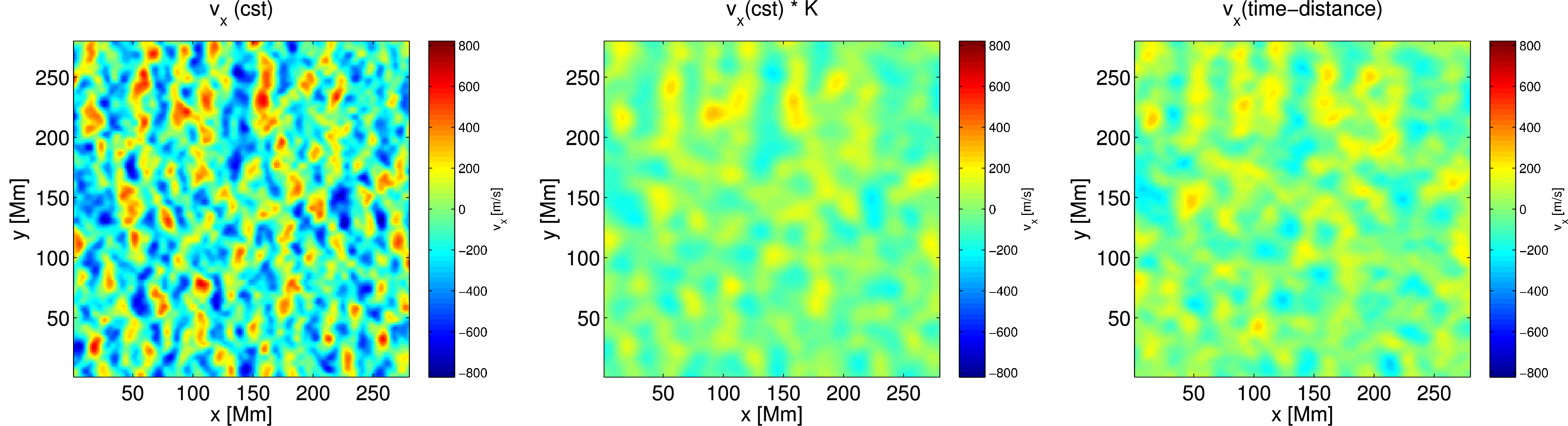}\\
\includegraphics[width=0.9\textwidth]{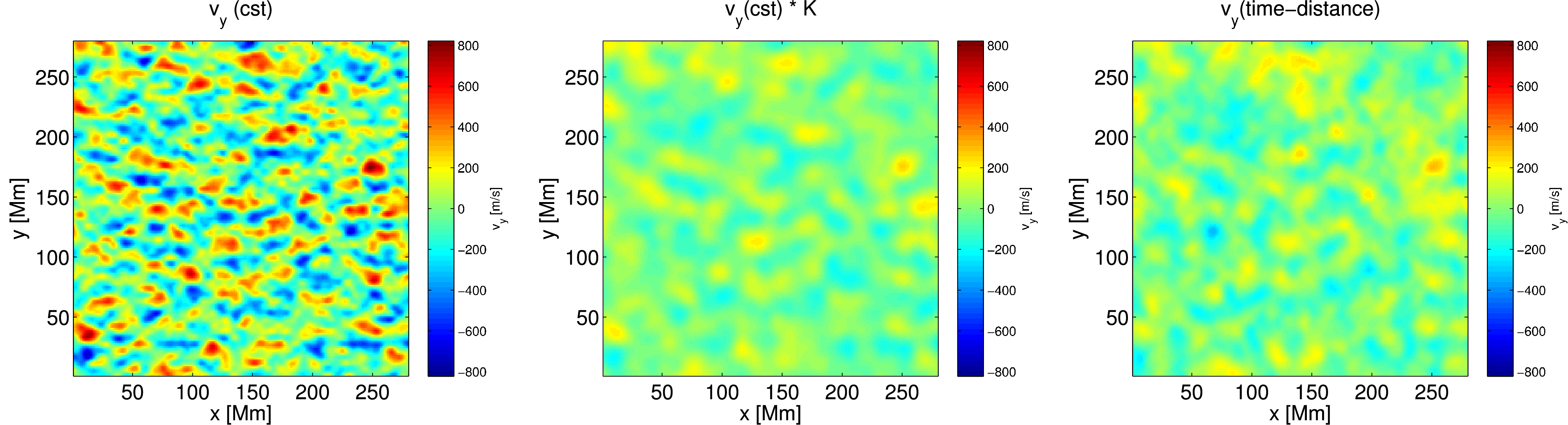}
\caption{Example maps of horizontal flow velocities ($v_x$ upper row and $v_y$ bottom row) measured from datacube No.~3. Displayed are (from left) CST-tracked velocity, granule motion velocity convolved with the inversion averaging kernel, estimate of the flow component from the inversion.}
\label{fig:maps}
\end{figure}

The co-aligned flow maps may be compared directly on a pixel-to-pixel basis. Both methods aim to measure plasma motions on the surface and should ideally match within the noise levels. That actually is almost true. Results coming from both methods have a different effective resolution. The CST flow estimates were computed using the full-resolution HMI intensitygrams and then binned seven times, thus the effective resolution is 3.5". The flow estimates from the time--distance inversion are obtained with the averaging kernel with $FWHM_h=10$~Mm, thus having an effective resolution of 13.7", much coarser. Thus, the flow maps from CST pipeline must naturally contain flows on smaller scales, which is clearly seen e.g. in Fig.~\ref{fig:maps} when comparing the left and right columns. To make flow maps from both methods comparable we convolved the velocity field from the CST code with a cut through the averaging kernel $\cK^\alpha_\alpha$ taken at $z=0$. The smoothed components of the CST flow are also displayed in Fig.~\ref{fig:maps} in the middle column.

Pixel-to-pixel statistics for all four datacubes are summarised in Table~\ref{tab:comparisons}. It gives the correlation coefficient ${\rm CC}$ for both horizontal components (for a curious reader, the correlation coefficient with non-smoothed flow estimate from CST is given in parentheses).

\begin{figure}
\includegraphics[width=0.49\textwidth]{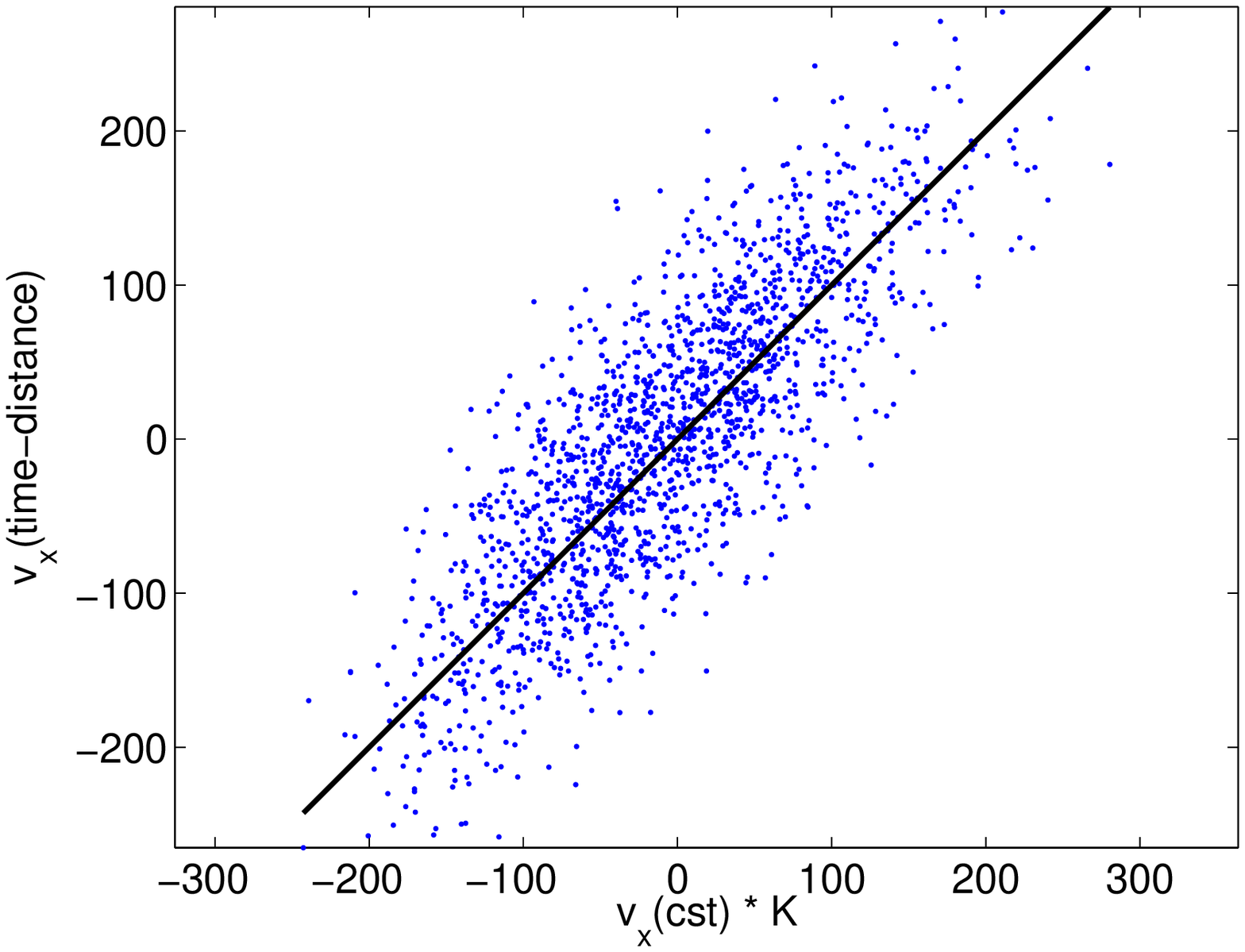}
\includegraphics[width=0.49\textwidth]{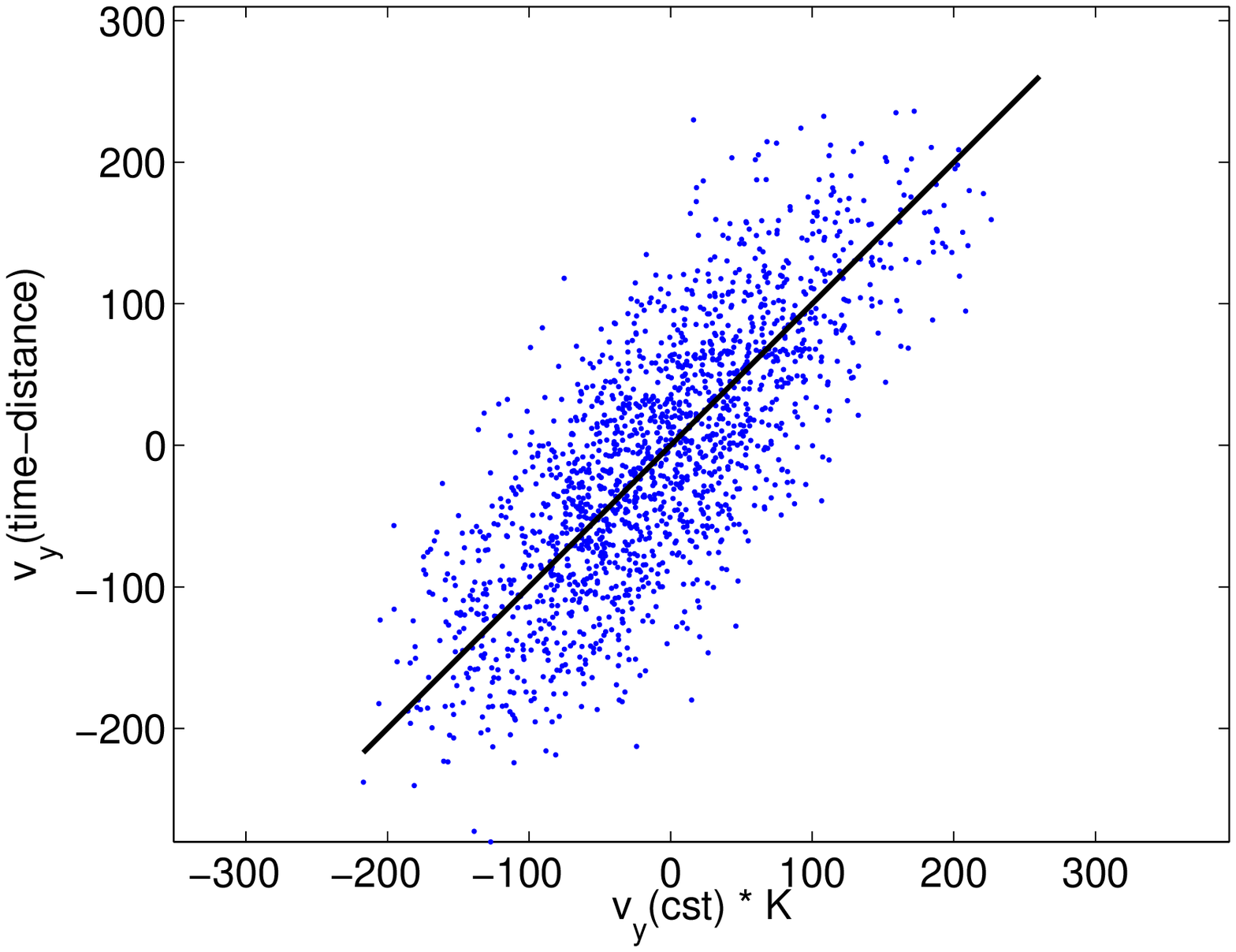}
\caption{Direct comparison of values of flow estimates from the two methods for independent points in the field of view. The line with slope of unity represents the location where all points should lie in case both flow estimates are identical. }
\label{fig:scatter}
\end{figure}

The comparison by eye from Fig.~\ref{fig:maps} reveals similarities but also some differences. To make the comparison more straightforward, we also plot values coming from both methods against each other in Fig.~\ref{fig:scatter}. The slope of the best least-squares fit through the data points using random errors in both variables \citep{2004AJF..72...367Y} is also given in Table~\ref{tab:comparisons}. The values measured by seismology are systematically smaller (by 3--5\%) than those measured by granule tracking. Such discrepancy can easily be explained by additional averaging of helioseismic flow estimates in the vertical direction. When assuming that the flow magnitudes peaks just at the surface and its magnitude drops rapidly in larger heights and decreases also in depth, the smoothing with the averaging kernels leads naturally to lower amplitudes. Possible large-magnitude (almost 1~k\mps) horizontal flows at depths around $1.6$~Mm suggested by \cite{2012SoPh..tmp..136D} do not affect this conclusion, because as pointed out by \cite{2012ApJ...759L..29S}, in inversion regularised strongly about the noise term, which is also the case here, the possible large-amplitude flows are largely smeared by the extended side-lobes of the averaging kernel. The spread of points seen in Fig.~\ref{fig:scatter} seems to be very consistent with the levels of random errors estimated for both methods.

Additional source of differences originates from the fact that both methods may not sample (or sample differently) the horizontal velocities of the same layers: granules move because their convective flux moves so the sampling function (vertical extend of the averaging kernel) of granules is the shape of their convective flux. On the other hand $f$-mode has another vertical profile which leads to another vertical sampling of the horizontal flow.

The location of the points, where the difference of the flow components measured by the two methods is significant (i.e., the speed values are above their respective noise levels for both methods and the difference is higher than the threshold discussed further), is random in the field of view -- these points do not correlate with the respective flow estimates. The threshold was determined from a simple model of the expected variance of the difference of two random variables, i.e., $\sigma_{\rm difference}=\sqrt{\sigma_{\rm cst}^2+\sigma_{\rm td}^2}$. The value of the theoretical threshold (55~\mps) is close to the variance of 62~\mps{} determined from studying the histogram of the differences of flow estimates from the two methods. The use of the latter value does not change the previous statement.

The comparison presented here serves as indication, that the helioseismic pipelines provide reasonable estimates of flows. It turns out that by directly comparing the velocity fields coming from granule tracking on the surface, we do not see any significant systematic errors.

\section{Vertical flow}

\begin{figure}
\includegraphics[width=0.49\textwidth]{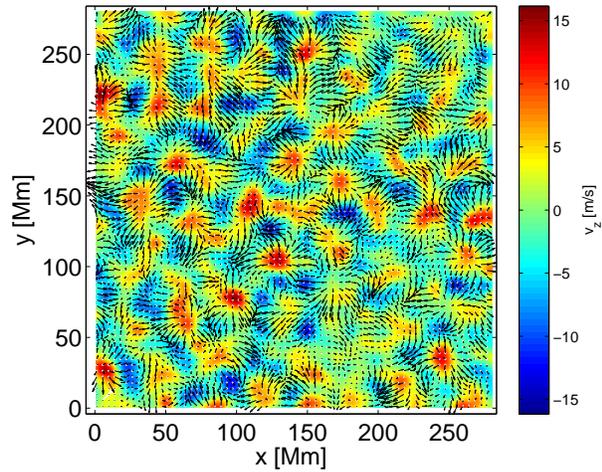}
\caption{An example of the full-vector flow map, horizontal components are displayed by arrows (reference arrow indicates flow 250~\mps), vertical component is presented by colours.}
\label{fig:fullvector}
\end{figure}

The helioseismic inversion allows us to measure the relative weak vertical flow at the surface \citep[thanks the to possibility to minimise the crosstalk;][]{Svanda2011}. The vertical velocity is usually not evaluated from helioseismic inversion and is usually modelled from the horizontal components assuming the mass conservation. In our case all three flow components are inverted for independently (an example of the full flow map displaying horizontal components by arrows and vertical by colour is displayed in Fig.~\ref{fig:fullvector}).

To test qualitatively the sensibility of the vertical flow inversion, we utilise the comparison of the vertical flow with the divergence of the horizontal flow, both measured by time--distance helioseismology. The averaging kernels are almost identical (see Fig.~\ref{fig:akerns}), thus the flow components are inverted for consistently. One would expect (due to the expected mass conservation of plasma flows and negligible horizontal variations in density) that divergent regions in the horizontal flow correspond to upflows in the vertical flow. Such behaviour is visible by eye in Fig.~\ref{fig:fullvector}. Quantitatively the correspondence can be evaluated by means of correlation coefficient between $v_z$ and $\partial_x v_x+\partial_y v_y$, which is summarised in Tab.~\ref{tab:vh_vs_vz}. Correlation is positive and significant. The correlation coefficient is expected to be higher, however one has to bear in mind that the horizontal and vertical flow components have different signal-to-noise ratios. That is estimated as ${\rm RMS}(v_\alpha)/\sigma_\alpha$ and has values 4.0 for horizontal, 1.1 for vertical components. Consequently, the signal-to-noise ratio of the horizontal divergence has value of 1.6. Thus one cannot expect perfect correspondence.

\subsection{Comparison with surface Dopplergram}

An additional surface measurement of the plasma flow exists, which can also be used to independently verify the results of the time--distance inversion for flows. It is the direct measurement of the Doppler component of velocity $v_{\rm d}$, high-cadence series of which is used to measure the travel times. Such comparison was already done in the past e.g. by \cite{2000JApA...21..339G}. In order to compare with the Doppler component, one has to project the velocity vector in the Cartesian system $(v_x,v_y,v_z)$ to line-of-sight component $v_{\rm los}$ and two complementary transversal components $(v_\xi, v_\zeta)$ using the heliographic coordinates of each point of the flow map which are known from the definition of the Postel projection. The transformation equations are given in a matrix form for illustration.

\begin{equation}
\left[ \begin{array}{c} v_\xi(\bvec{r}) \\ v_\zeta(\bvec{r}) \\ v_{\rm los}(\bvec{r}) \end{array} \right] =
\left[ \begin{array}{ccc} 1 & 0 & 0 \\ 0 & \cos\varphi(\bvec{r}) & -\sin\varphi(\bvec{r}) \\ 0 & \sin\varphi(\bvec{r}) & \cos\varphi(\bvec{r}) \end{array} \right]
\left[ \begin{array}{ccc}
\cos\vartheta(\bvec{r}) & 0 & -\sin\vartheta(\bvec{r}) \\ 0 & 1 & 0 \\ \sin\vartheta(\bvec{r}) & 0 & \cos\vartheta(\bvec{r})
\end{array} \right]
\left[ \begin{array}{c} v_x(\bvec{r}) \\ v_y(\bvec{r}) \\ v_z(\bvec{r}) \end{array} \right],
\end{equation}
where $\vartheta(\bvec{r})=l(\bvec{r})-l_0$ and $\varphi(\bvec{r})=b(\bvec{r})-b_0$ are the heliocentric coordinates, computed from the Carrington coordinates $l$ and $b$.

\begin{figure}
\includegraphics[width=0.9\textwidth]{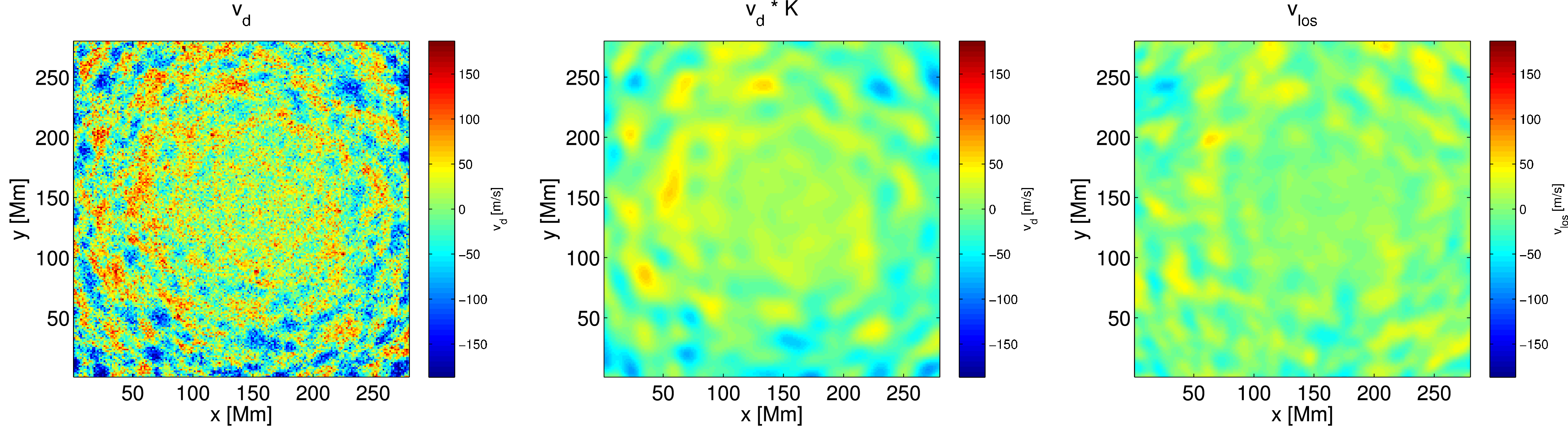}
\caption{Comparison of the measured Doppler component of the velocity (left), the measured Dopplergram convolved with the averaging kernel (middle), and the time--distance vector flow estimate projected to the line-of-sight (right). One sees a striking correspondence between the middle and right panels. }
\label{fig:doppler}
\end{figure}

The reference Dopplergram for each of four datacubes was derived as a temporal average of all high-cadence Dopplergrams over 12 hours. As in the case of the comparison to the horizontal flow derived from the granule tracking discussed in Section~\ref{sect:cst_vs_td}, the direct Dopplergram is obtained with a different effective spatial resolution. Thus it must contain information on smaller scales than the results from the flow inversion, which acts as noise and naturally decreases the correlation with the projected line-of-sight velocity. Thus, as in Section~\ref{sect:cst_vs_td}, we convolved the averaged Dopplergram with the horizontal section of the inversion averaging kernel. An example of three comparable maps (Dopplergram, spatially averaged Dopplergram, and a line-of-sight velocity projection) is displayed in Fig.~\ref{fig:doppler}. The Pearson correlation coefficient between the latter two for each of the four datacubes investigated is given in Table~\ref{tab:vh_vs_vz}. The correlation is positive and significant. Note that \cite{2000JApA...21..339G} mention slightly higher correlation between the Dopplergram and a line-of-sight projection (correlation coefficient of 0.7). The lower correlation coefficient in our study originate from the region at the disc centre, where the weak (and quite noisy) vertical component $v_z$ of the flow dominates the projected line-of-sight velocity. When the $v_z$ is excluded (set to zero) from the line-of-sight velocity projection, the correlation coefficient is by 0.05 higher for all four datacubes. When the centre-of-the-disc region is omitted (central 70 Mm), the correlation coefficient increases by additional 0.03.

\subsection{Density scale height}

From a direct comparison between the horizontal and vertical components of the flow we estimated the density scale height. Let us start from separating the density $\rho$ and velocity vector $\vv$ into time-averaged components $\rhobar$ and $\vvb$ and fluctuations $\rho'$ and $\vv'$. Thus the continuity equation now reads

\begin{equation}
\frac{\partial \rho'}{\partial t} + \nabla \cdot \left\{ (\rhobar+\rho')(\vvb+\vv') \right\} =0.
\label{eq:continuity}
\end{equation}

By taking the time average of Eq.~(\ref{eq:continuity}) and assuming that $\overline{\rho'}=0$ and $\overline{\vv'}=0$ we obtain

\begin{equation}
\vvb\cdot\nabla\ln\rhobar+\nabla\cdot\vvb_h +
\dz{\vbar_z}+ \frac{1}{\rhobar} \nabla\cdot\overline{\rho'\vv'} = 0
\end{equation}
where $\vvb_h$ is the average horizontal velocity. Finally we get

\begin{equation}
\vbar_z\dz{\ln\rhobar}+\nabla\cdot\vvb_h =
-\dz{\vbar_z}-\frac{1}{\rhobar} \nabla\cdot\overline{\rho'\vv'}
\label{eq:final_for_fit}
\end{equation}
where we assumed that the averaged density depends only on depth. We see that if the right-hand side is negligible, then the horizontal divergence and the vertical velocity are correlated. Moreover, the slope of the correlation gives the scale height of the density $H_\rho=-\left[ \partial \ln \rhobar / \partial z\right]^{-1}$. Since horizontal velocities and vertical ones are from the same layers, the correlation slope gives a weighted average of the density scale height at the sampled depth.

% \begin{figure}
% \includegraphics[width=0.49\textwidth]{scale_height.eps}

% \caption{Vertical velocity versus horizontal divergence for flows from time--distance inversion. The lines show a 1-$\sigma$ range of the least-squares fit when errors in both plotted variables are taken into account. The slope of the fit roughly estimates the density scale height near the surface.}
% \label{fig:scaleheight}
% \end{figure}
%
% The combined plot for all four datacubes investigated is displayed in Fig.~\ref{fig:scaleheight}.

Doing the exercise, the least-squares fit of the data by a linear law using the known error levels in both variables \citep{2004AJF..72...367Y} gives an estimate for the density scale height in the near-surface layers, which we find around $180$~km. Such a value is of the same order as the 125~km found by \cite{1994SoPh..154....1N} from a study of the supergranulation. By simply taking the density scale height from Model S \citep{1996Sci...272.1286C} and integrating it over the inversion averaging kernel, we get a value of 535~km, which is three times larger than the scale height determined from the inversion. However, one should note that the density scale height determinantion is a sanity check of inversion consistency. Additionally, due to the presence of additional terms in Eq.~(\ref{eq:final_for_fit}), one cannot expect a better than an order of magnitude agreement with the solar model.

\section{Conclusion}

We measured surface horizontal flows using both coherent structure tracking and 3-D time-distance inversions. The measurements from the two techniques are in good agreement. Our future goals will focus on studies of plasma dynamics especially on studies of solar convection \citep[see][]{2012PNAS..10911896G,2012PNAS..10911928H}.

% This is an important step towards the scientific goals exploiting the properties of the convection in the upper layers of the solar convection zone. The recent ``convection crisis'' \citep[see][]{2012PNAS..10911896G,2012PNAS..10911928H} clearly shows that our knowledge of the properties of the solar convection below the solar surface is rather poor. It seems now that only helioseismology, performed using validated methods with minimised biases may answer the questions about the sub-surface convection.

\acknowledgements M.\v{S} is supported by the Czech Science Foundation (grant P209/12/P568). This study was supported by the European Research Council under the European Community's Seventh Framework Programme (FP7/2007--2013)/ERC grant agreement \#210949, ``Seismic Imaging of the Solar Interior'', to PI L. Gizon (Milestone \#5). L. G. acknowledges research funding by Deutsche Forschungsgemeinschaft (DFG) under grant SFB 963/1 ``Astrophysical flow instabilities and turbulence'' (Project A1). This work utilised the resources and helioseismic products provided by the German Data Center for the Solar Dynamics Observatory (GDC-SDO), which is hosted by the MPS in Katlenburg-Lindau, Germany, and funded by the German Aerospace Center (DLR). The data were kindly provided by the HMI consortium. The HMI project is supported by NASA contract NAS5-02139. Tato pr\'ace vznikla s podporou na dlouhodob\'y koncep\v{c}n\'\i{} rozvoj v\'yzkumn\'e organizace RVO:67985815 a v\'yzkumn\'eho z\'am\v{e}ru MSM0021620860.

% \bibliographystyle{apj}
% \bibliography{inversions}

%%%%%%%%%%%%%%%%%%%%%%%%%%%%%%%%%%%%%%%%%%%%%%%%%%%%%%%%%%% FIGURES and TABLES %%%%%%%%%%%%%%%%%%%%%%%%%%%%%%%%%%%%%%%%%%%%%%%%%%%%%%%%%%%%%%%
\clearpage

\begin{table}
\begin{tabular}{l|llll}
\hline
Cube No. & Time range & $b_0$ & $l_0$ & No. missing frames\\
\hline
1 & 12 May 2012 00:01:30--12:01:30 TAI & -3.00 & 324.94 & 12\\
2 & 12 May 2012 12:01:30--24:01:30 TAI & -3.00 & 318.33 & 6\\
3 & 13 May 2012 00:01:30--12:01:30 TAI & -3.00 & 311.72 & 6\\
4 & 13 May 2012 12:01:30--24:01:30 TAI & -3.00 & 305.11 & 6\\
\hline
\end{tabular}
\caption{Datacubes used in the comparative analysis: temporal coverage and Carrington coordinates ($l_0$ for longitude and $b_0$ for latitude) of the central point. Also number of missing frames (out of 961) are given for a curious reader.}
\label{tab:cubes}
\end{table}

\begin{table}
\begin{tabular}{l|cccc}
\hline
Cube No. & ${\rm CC}(v_{x,{\rm cst}},v_{x,{\rm td}})$ & ${\rm CC}(v_{y,{\rm cst}},v_{y,{\rm td}})$ & ${\rm slope}(v_{x,{\rm cst}},v_{x,{\rm td}})$ & ${\rm slope}(v_{y,{\rm cst}},v_{y,{\rm td}})$ \\
\hline
1 & 0.78 (0.60) & 0.75 (0.57) & 0.97 & 0.94\\
2 & 0.78 (0.59) & 0.73 (0.56) & 0.94 & 0.89\\
3 & 0.82 (0.63) & 0.76 (0.56) & 0.96 & 0.93\\
4 & 0.79 (0.61) & 0.73 (0.56) & 0.95 & 0.85\\
\hline
\end{tabular}
\caption{Statistical values for comparison of results of the two methods. We give the correlation coefficient ${\rm CC}$ of the horizontal flow components and the slope of the least-squares fit to the flow estimates obtained by the two methods.}
\label{tab:comparisons}
\end{table}

\begin{table}
\begin{tabular}{l|cc}
\hline
Cube No. & ${\rm CC}(v_{z},\partial_x v_x+\partial_y v_y)$ & ${\rm CC}(v_{\rm f},v_{\rm los})$\\
\hline
1 & 0.50 & 0.59\\
2 & 0.47 & 0.57\\
3 & 0.49 & 0.64 \\
4 & 0.52 & 0.51 \\
\end{tabular}
\caption{Correlation between vertical flow $v_z$ and the horizontal divergence of the horizontal flow and correlation between the measured Doppler velocity and the vector velocity field projected into the line of sight.}
\label{tab:vh_vs_vz}
\end{table}

\end{document}